\documentclass[aps,prl,reprint,onecolumn,amsmath,amssymb,superscriptaddress,]{revtex4-1}

\usepackage{graphicx}
\usepackage{dcolumn}
\usepackage{bm}
\usepackage{xcolor}
\newcommand{\LiFeO}{(Li$_{0.8}$Fe$_{0.2}$)ODFeSe}

\begin{document}

\preprint{APS/123-QED}

\title{Polarized neutron scattering studies of magnetic excitations in iron-selenide superconductor (Li$_{0.8}$Fe$_{0.2}$)ODFeSe ($T_c$ = 41 K)}

\author{Die Hu}
\affiliation{State Key Laboratory of Surface Physics and Department of Physics, Fudan University, Shanghai 200433, China}

\author{Yu Feng}
\affiliation{State Key Laboratory of Surface Physics and Department of Physics, Fudan University, Shanghai 200433, China}

\author{Jitae T. Park}
\affiliation{Heinz Maier-Leibnitz Zentrum (MLZ), Technische Universit\"at M\"unchen, D-85748 Garching, Germany}

\author{Hongliang Wo}
\affiliation{State Key Laboratory of Surface Physics and Department of Physics, Fudan University, Shanghai 200433, China}

\author{Qisi Wang}
\affiliation{State Key Laboratory of Surface Physics and Department of Physics, Fudan University, Shanghai 200433, China}

\author{Fr$\rm{\acute{e}}$d$\rm{\acute{e}}$ric Bourdarot}
\affiliation{Universit$\acute{e}$ Grenoble Alpes, CEA, INAC, MEM MDN, F-38000 Grenoble, France}

\author{Alexandre Ivanov}
\affiliation{Institut Laue-Langevin, 71 Avenue des Martyrs CS 20156, 38042 Grenoble Cedex 9, France}

\author{Jun Zhao}
\email{zhaoj@fudan.edu.cn}
\affiliation{State Key Laboratory of Surface Physics and Department of Physics, Fudan University, Shanghai 200433, China}
\affiliation{Institute for Nanoelectronic Devices and Quantum Computing, Fudan University, Shanghai 200433, China}
\affiliation{Shanghai Qi Zhi Institute, Shanghai, 200232, China}
\affiliation{Shanghai Research Center for Quantum Sciences, Shanghai 201315, China}

\begin{abstract}
{
We report polarized neutron scattering measurements of the low energy spin fluctuations of the iron-selenide superconductor Li$_{0.8}$Fe$_{0.2}$ODFeSe below and above its superconducting transition temperature $T_c=41$ K. Our experiments confirmed that the resonance mode near 21 meV is magnetic. Moreover, the spin excitations are essentially isotropic in spin space at 5$\leq E\leq$ 29 meV in the superconducting and normal states. Our results suggest that the resonance mode in iron-based superconductors becomes isotropic when the influence of spin-orbit coupling and magnetic/nematic order is minimized, similar to those observed in cuprate superconductors.
}

\end{abstract}
\maketitle

 The spin resonance mode is the most prominent feature in the spin fluctuation spectrum in numerous unconventional superconductors, including cuprate, heavy fermion and iron-based superconductors\cite{Rossat-Mignod, Scalapino2012RMP,Steglich2016RPR, Matthias2006AP,Dai2015RMP}. The resonance mode occurs below the superconducting transition temperature ($T_c$) and exhibits a temperature dependence similar to the superconducting order parameter. The energy of resonance mode also scales with $T_c$ among different classes of unconventional superconductors\cite{Matthias2006AP,Dai2015RMP}. Therefore, the resonance mode has been considered as the hallmark of unconventional superconductivity.

 It has been proposed that the resonance mode is a spin-triplet exciton of the singlet Cooper pairs with a sign-reversed superconducting gap function\cite{Matthias2006AP}. The observation of a spin-space isotropic resonance mode at the antiferromagnetic wavevector in hole-\cite{Headings2011PRB} and electron-doped \cite{Jun2011NP} cuprate superconductors is consistent with this picture with a $d$-wave pairing gap function. In most iron pnictide superconductors, the resonance modes appear near the stripe antiferromagnetic wavevector, which is close to the wavevector connecting the electron pockets at the Brillouin zone corner and the hole pockets at the zone center\cite{Christianson2008N,Lumsden2009PRL,Chi2009PRL,Inosov2010NP, Zhang2013PRL,Zhao2013PRL}. This was explained as a spin exciton arising from the sign-reversed $s$-wave pairing between the nested electron and hole pockets\cite{Dai2015RMP}.

However, the situation in iron-based superconductors seems more complicated than expected, as recent polarized neutron scattering measurements revealed that the resonance modes in most underdoped/optimally doped and even some overdoped iron pnictides are anisotropic in spin space\cite{Lipscombe2010PRB, Li2017PRB, WaBer2017SR, Zhang2014PRB,Steffens2013PRL,Qureshi2014PRB}, which challenged the general notion that the spin resonance is an isotropic triplet spin-exciton. One possible explanation for the anisotropy of the spin exciton could be related to adjacent magnetic/nematic order phases with moderate spin-orbit coupling\cite{Day2018PRL,Daniel2018PRL}. Alternatively, the resonance mode was attributed to magnonlike excitations similar to those of the magnetically ordered parent compounds, where spin-orbit coupling has been shown to lead to anisotropic magnon gaps\cite{Qureshi2012PRB,npj2019}.

\begin{figure}
\centering
\includegraphics[width=8.6 cm]{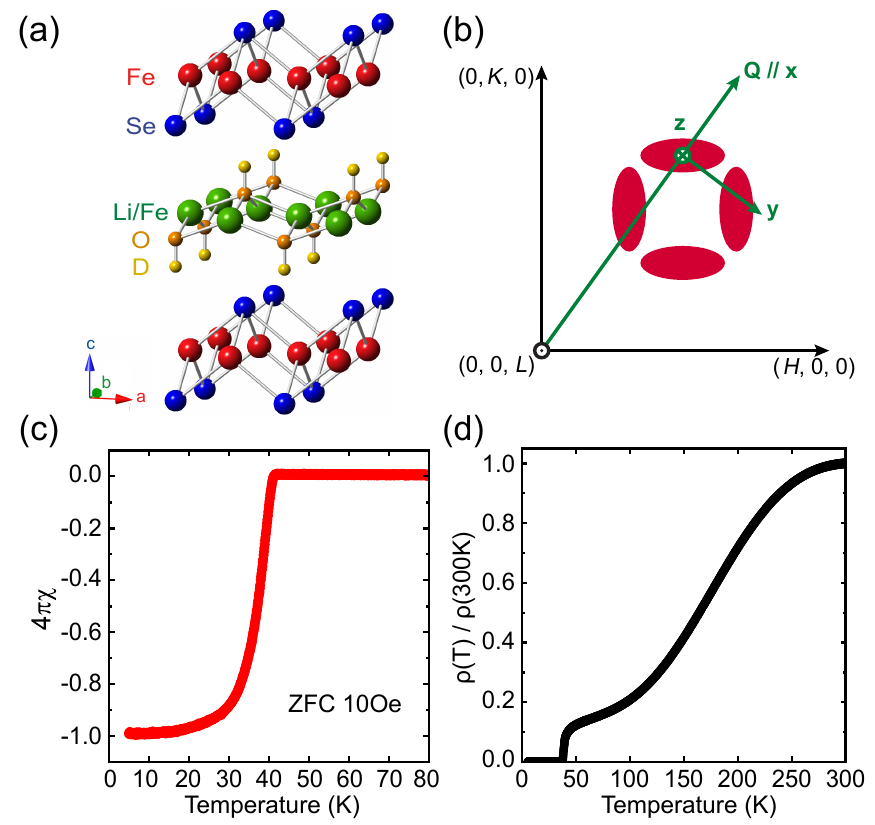}
\centering
\caption{\label{fig1}
$\mathrm{\bf{a}}$ Crystal structure of (Li$_{0.8}$Fe$_{0.2}$)ODFeSe. FeSe layers and (Li$_{0.8}$Fe$_{0.2}$)OD layers stack alternately. $\mathrm{\bf{b}}$ A schematic of the scattering geometry and definition of polarization in our experiment. The red ellipses represent the spin resonance of (Li$_{0.8}$Fe$_{0.2}$)ODFeSe in the momentum space (1-Fe unit cell). $\mathrm{\bf{c}}$ Magnetic susceptibility in the zero-field-cooled (ZFC) measurement in a magnetic field of H = 10 Oe, indicating $100\%$ exclusion of the external magnetic field. $\mathrm{\bf{d}}$ Resistance as a function of temperature of (Li$_{0.8}$Fe$_{0.2}$)ODFeSe single crystal. A sharp superconducting transition is observed at $T_c$ = 41 K. }
\end{figure}

As yet all previous polarized neutron scattering studies have been focused on the iron-based superconductors with both electron and hole pockets\cite{Lipscombe2010PRB, Li2017PRB, WaBer2017SR, Zhang2014PRB,Steffens2013PRL,Qureshi2014PRB,Ma2017PRX,Babkevich2011PRB}. The resonance modes have also been observed in electron-doped iron-selenide superconductors with no hole pockets\cite{ParkPRL2011, Boothroyd, MWangPRB2012, inosov2012EPL, Wang2016PRL}, but the spin anisotropy in these materials remains to be explored. The recently synthesized electron-doped iron-selenide superconductor Li$_{0.8}$Fe$_{0.2}$OHFeSe ($T_c = 41$ K) has attracted great interest as it has no hole pockets at the zone center but has a relatively high $T_c$ (Fig. 1(a)) \cite{Lu2015NM,Pachmayr2015ACIE,Zhao2016NC,Niu2015PRB}.  Li$_{0.8}$Fe$_{0.2}$OHFeSe is ideal for polarized neutron scattering measurements as it is hitherto the only electron-doped iron-selenide superconductor that phase-pure and large-size single-crystalline samples can be grown. Our previous unpolarized neutron scattering measurements revealed that the resonance mode in Li$_{0.8}$Fe$_{0.2}$ODFeSe exhibits a nearly ring-shaped structure surrounding ($\pi$, $\pi$), which is far away from the stripe magnetic wavevector ($\pi$, 0)\cite{Pan2017NC}. This, together with the facts that Li$_{0.8}$Fe$_{0.2}$ODFeSe does not exhibit magnetic or nematic order associated with the FeSe plane, suggests that the influence of the remnant static or quasi-static magnetic/nematic order should be minimized. Therefore, Li$_{0.8}$Fe$_{0.2}$ODFeSe is ideally suitable for revealing the intrinsic properties of the resonance mode and their interplay with superconductivity.

In this paper, we use polarized neutron scattering to study the low energy spin fluctuations of single crystalline Li$_{0.8}$Fe$_{0.2}$ODFeSe ($T_c=41$ K) below and above $T_c$. Our experiments confirmed that the resonance mode near 21 meV is magnetic. Moreover, the spin excitations are isotropic in spin space below 29 meV within the error of the measurements in the superconducting and normal states. These results are in agreement with the scenario that the resonance mode is an isotropic spin-triplet exciton.

\begin{figure}
\includegraphics[width=9.5cm]{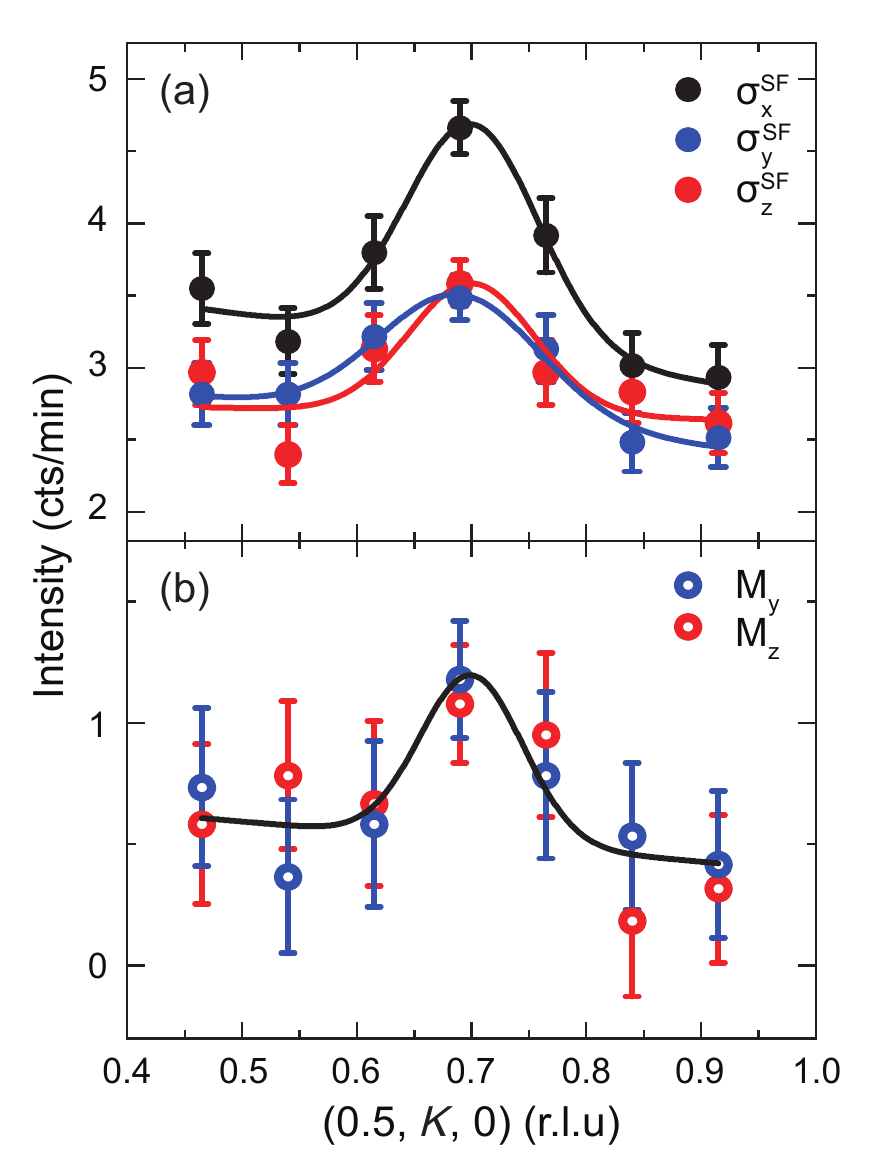}
\caption{\label{fig2}
$\mathrm{
\bf{a}}$ Constant energy scans of $\sigma_x^{SF}$, $\sigma_y^{SF}$ and $\sigma_z^{SF}$ through the resonance wavevector (0.5, 0.69, 0) along the K direction for $E$ = 21 meV at 2 K. $\mathrm{\bf{b}}$  M$_y$ = $\sigma_x^{SF}$ - $\sigma_y^{SF}$ and M$_z$ = $\sigma_x^{SF}$ - $\sigma_z^{SF}$ determined from data in $\mathrm{\bf{a}}$. Solid lines were fits using a Gaussian function on a linear background. No background correction was performed for all data presented.}
\end{figure}

Li$_{0.8}$Fe$_{0.2}$ODFeSe single crystals were synthesized by the hydrothermal ion-exchange method\cite{Dong2015PRB}. K$_x$Fe$_{2-y}$Se$_2$ single crystals were first grown using the flux method\cite{Wang2016PRL}. The starting materials were high-purity K pieces, Fe powder, and Se pieces with the nominal composition K$_{0.8}$Fe$_2$Se$_2$. The reaction mixtures were sealed in evacuated double-walled quartz ampoules and heated to 1030$^{\circ}$C, and then slowly cooled to 730$^{\circ}$C within 100 hours before shutting off the heating power of the furnace. The obtained K$_x$Fe$_{2-y}$Se$_2$ single crystals show a superconducting transition temperature of $\sim31$ K. Around 0.6 g of K$_x$Fe$_{2-y}$Se$_2$ single crystals were cut into thin plates with the widest faces perpendicular to the $c$ direction before mixing with Fe powder (0.3 g), Li$_2$O (0.01 mol), selenourea (0.0075 mol) and lithium deuteroxide solution ($\sim$ 10 ml) in a Teflon-line steel autoclave. The autoclave was tightly sealed and heated to 120$^{\circ}$C for 5 days, then naturally cooled down to room temperature. The obtained Li$_{0.8}$Fe$_{0.2}$ODFeSe crystals were washed with deionized water using a B\"uchner flask with circulating pump. We note that deuteroxide was used during the growth to reduce the incoherent scattering from hydrogen atoms for the neutron scattering measurements.

Magnetic susceptibility and in-plane resistivity measurements on Li$_{0.8}$Fe$_{0.2}$ODFeSe single crystals from the same batches as the neutron scattering crystals show a sharp superconducting transition temperature of 41 K [Figs. 1(c) and 1(d)], which indicates the high quality of the crystals. Refinement results from powder x-ray diffraction on grind single crystals are consistent with previous reports \cite{Pan2017NC,Lu2015NM}.

\begin{figure*}
\centering
\includegraphics[width=17.5cm]{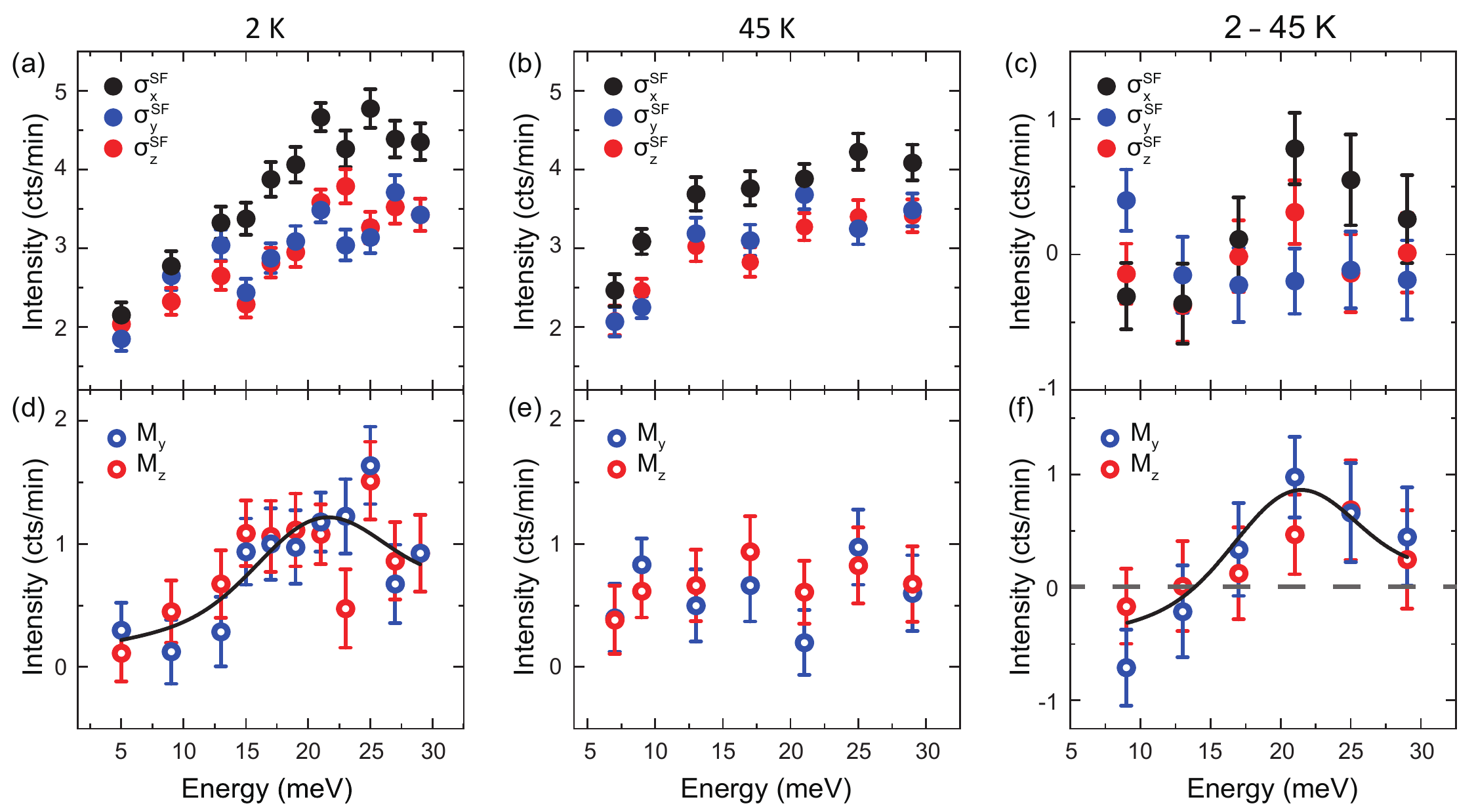}
\centering
\caption{
 Energy scans of $\sigma_x^{SF}$, $\sigma_y^{SF}$, and $\sigma_z^{SF}$ at the resonance wavevector Q = (0.5, 0.69, 0) at 2 K (a), 45 K (c) and the intensity difference between 2 K and 45 K (d) . $\mathrm{\bf{d}}$ $\mathrm{\bf{e}}$ $\mathrm{\bf{f}}$ Magnetic components  M$_y$ and M$_z$ determined from $\mathrm{\bf{a}}$ $\mathrm{\bf{b}}$ $\mathrm{\bf{c}}$, respectively. Solid lines are a guide to the eye. }
\end{figure*}

Polarized neutron scattering experiment was carried out on IN22, CRG-CEA thermal triple-axis spectrometer at Institut Laue-Langevin (ILL) in Grenoble, France. The final neutron wavevector was fixed at $k_f$ = 2.664 \AA$^{-1}$. The Heusler monochromator and analyzer were used for the experiment. We coaligned $\sim$ 4 g \LiFeO~ single crystals in the ($H$, $K$, 0) scattering plane with a mosaicity of $\sim3.4^{\circ}$. The flipping ratio, determined on the (100) nuclear Bragg reflection by the leakage of non-spin-flip (NSF) channel into spin-flip (SF) channel, approximates to $R$ = $\sigma_{Bragg}^{NSF}$ / $\sigma_{Bragg}^{SF}$ $\approx$ 10. No corrections were done for the instrumental polarization leakage. In this experiment we defined the neutron polarization directions as $x, y, z$ with $x$ parallel to $\bf{Q}$, $y$ perpendicular to $\mathbf{Q}$ and in the scattering plane, and $z$ perpendicular to the scattering plane. In Fig. 1(b), the green arrows indicate the direction of neutron polarization and the red ellipses represent scattering signals in reciprocal space in one-Fe unit cell in (Li$_{0.8}$Fe$_{0.2}$)ODFeSe adapted from ref. \cite{Pan2017NC}. In a neutron scattering process, only the scattering components perpendicular to the momentum transfer $\mathbf{Q}$ are probed. With current experimental geometry, the SF cross sections can be written as\cite{Squires1978}:
$$ \sigma_x^{SF}=M_y+M_z+BG $$
$$ \sigma_y^{SF}=M_z+BG  $$
$$ \sigma_z^{SF}=M_y+BG   \eqno{(1)} $$

In Eq. (1), $BG$ represents the background. $M_y = \sigma_x^{SF}-\sigma_z^{SF}$, $M_z = \sigma_x^{SF}-\sigma_y^{SF}$. When the magnetic excitation is isotropic in spin space, magnetic scattering should have the same intensity in different orientations, meaning that $M_y$ is equal to $M_z$ and that $\sigma_y^{SF}$ is equal to $\sigma_z^{SF}$.

Fig. 2(a) shows the constant energy scans at 21 meV and 2 K near (0.5, 0.69, 0), where the resonance mode was observed in our previous unpolarized neutron scattering experiments \cite{Pan2017NC}. It is shown that all three SF channels $\sigma_x^{SF}$, $\sigma_y^{SF}$, and $\sigma_z^{SF}$ display well-defined peaks, which confirms that the resonance mode is magnetic. We note that the peaks in $\sigma_y^{SF}$, and $\sigma_z^{SF}$ exhibit nearly the same peak intensity. This immediately implies that the magnetic excitation at this energy is isotropic. In fact, based on Eq. (1), we can extract $M_y$ and $M_z$ from Fig. 2(a), which displays equal counts within the statistical error.

 In order to determine the energy dependence of the spin excitation anisotropy, we measured the constant-$\bf{Q}$ scans for three SF cross sections $\sigma_x^{SF}$, $\sigma_y^{SF}$ and $\sigma_z^{SF}$ from 5 meV to 29 meV at $\bf{Q}$ = (0.5, 0.69, 0) in the superconducting ($T$ = 2 K) and normal ($T$ = 45 K) states.  As shown in Fig. 3(a), scattering signals in all three SF channels exhibit a peak feature near $21$ meV at 2 K, with $\sigma_x^{SF}$ displaying the highest intensity. This is consistent with the $\mathbf{Q}$-scans. On warming to 45 K above $T_c$, the peak at $\sim$ 21 meV weakened while the relative intensities among three different cross sections remain essentially unchanged. With all 3 SF cross sections determined we can extract $M_y$ and $M_z$ below and above $T_c$. It is found that $M_y$ and $M_z$ have roughly the same intensity from 5 meV to 29 meV below and above $T_c$. This corroborates our conclusion that the spin fluctuations are isotropic in spin space. To elucidate the interplay between magnetic excitations and superconductivity, we plot in Fig. 3(c) and 3(f) the energy dependence of the intensity difference between the superconducting and normal states, which again revealed an essentially isotropic resonance mode.

The anisotropy of the resonance mode in iron pnictides has been attributed to the influence of the nearby magnetic/nematic order phase, as the anisotropic resonance excitations are rather similar to the spin wave excitations with anisotropic magnon gaps in the magnetically ordered parent compounds\cite{Qureshi2012PRB}. This is supported by the fact that the resonance mode in overdoped BaFe$_{1.85}$Ni$_{0.15}$As$_2$ ($T_c=14$ K) becomes essentially isotropic in spin space\cite{Liu2012PRB}. On the other hand, recent polarized neutron scattering experiments have revealed anisotropic spin fluctuations even in overdoped Ba$_{0.5}$K$_{0.5}$Fe$_2$As$_2$ \cite{Qureshi2014PRB}, which suggests that the nearby magnetic/nematic order may not be the only reason causing the spin anisotropy. In fact, band structure calculations and measurements have revealed sizable band splitting due to spin-orbit coupling in many iron-based superconductors\cite{S.V.Borisenko2016NP}, which may induce anisotropic spin excitations even in overdoped samples. However, a close inspection of the band structure measurements and calculations show that the band splitting mainly occurs in the hole bands near the zone center, while the band splitting in the electron bands near the zone corner is less significant\cite{Day2018PRL,S.V.Borisenko2016NP,Liu2018PRB}. Since Li$_{0.8}$Fe$_{0.2}$ODFeSe has no hole pockets near the zone center and the low energy spin fluctuations are likely only related to the electron pockets near the zone corner, the influence of the spin-orbit coupling to the low energy spin fluctuations is minimized. This naturally explains the absence of the anisotropy of the low energy spin fluctuations in Li$_{0.8}$Fe$_{0.2}$ODFeSe.

To summarize, we present polarized neutron scattering studies of the low energy spin fluctuations in iron-selenide superconductor Li$_{0.8}$Fe$_{0.2}$ODFeSe with only electron Fermi pockets. Our data unambiguously demonstrate that the resonance mode near 21 meV is magnetic. Moreover, contrary to the many iron pnictide superconductors with anisotropic low energy spin fluctuations, the low energy spin excitations in Li$_{0.8}$Fe$_{0.2}$ODFeSe ($T_c = 41$ K) are isotropic in spin space within the accuracy of the measurements in both the superconducting and normal states. The diminished anisotropy of spin fluctuations is likely due to the absence of the static or quasi-static magnetic/nematic order and the weakened spin-orbit coupling effects on the electron pockets in Li$_{0.8}$Fe$_{0.2}$ODFeSe. The observed isotropic resonance mode in Li$_{0.8}$Fe$_{0.2}$ODFeSe is similar to those of cuprate superconductors, and is consistent with an isotropic spin-triplet exciton of the singlet Cooper pairs with a sign-reversed superconducting gap function.

This work was supported by the National Natural Science Foundation of China (Grant No.11874119), the Innovation Program of Shanghai Municipal Education Commission (Grant No. 2017-01-07-00-07-E00018), the National Key R\&D Program of the MOST of China (Grant No. 2016YFA0300203) and the Shanghai Municipal Science and Technology Major Project (Grant No.2019SHZDZX01).

\newpage
\section{Appendix}
\setcounter{figure}{0}
\renewcommand{\figurename}{FIG. A}

\begin{center}
\textbf{I. Powder X-ray diffraction of Li$_{0.8}$Fe$_{0.2}$ODFeSe}    
\end{center}

X-ray diffraction measurement at 300 K has been performed on the polycrystalline sample ground from the single crystals (Fig. A1). The refined lattice parameters \textit{a} = \textit{b} = 3.78639(24) \AA \quad and \textit{c} = 9.31217(48) \AA~  are consistent with previous reports \cite{Pan2017NC,Lu2015NM}.

\begin{figure}[h]
\centering
\includegraphics[width=12cm]{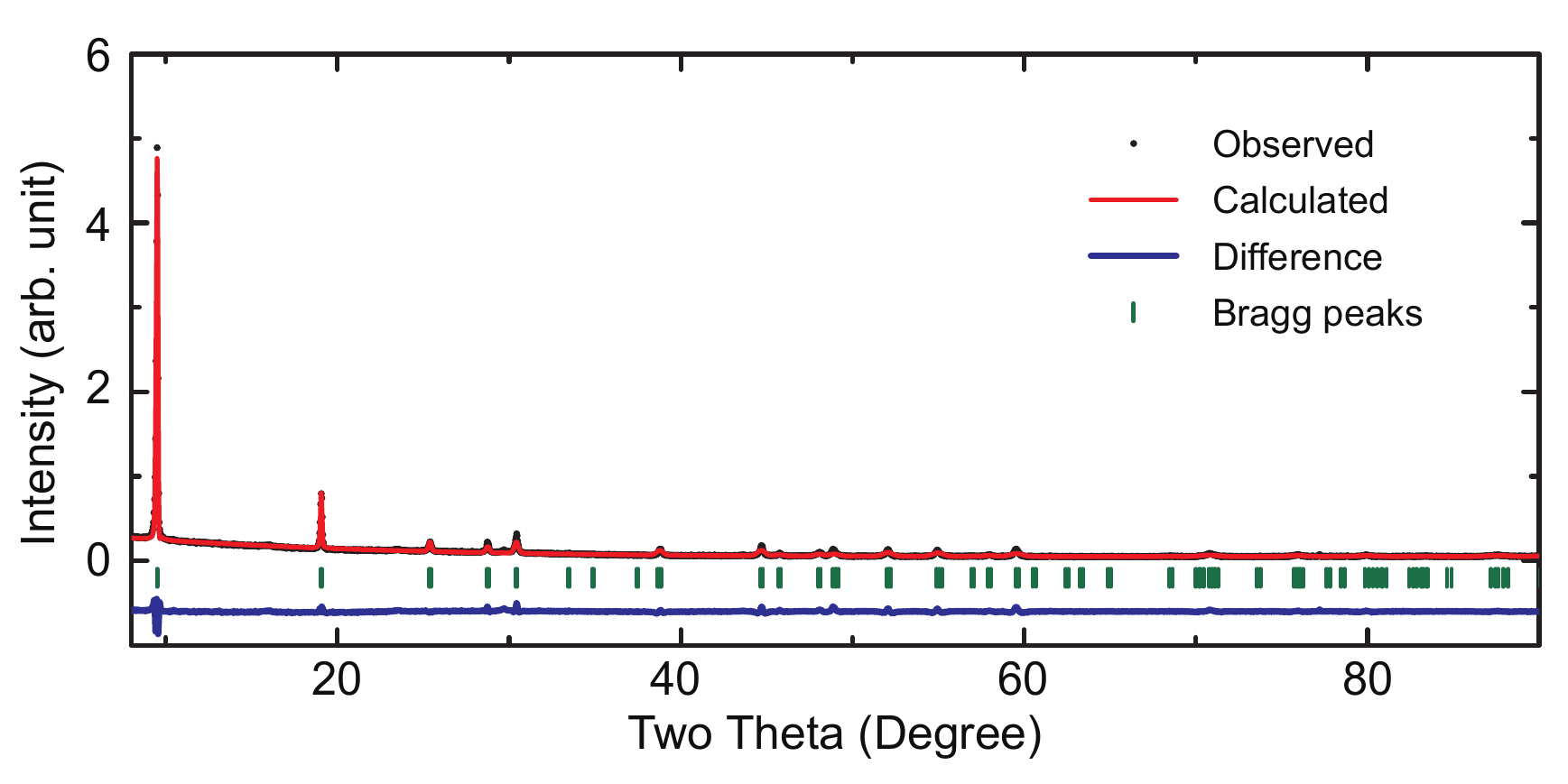}
\centering
\caption{
 Observed and calculated X-ray powder diffraction pattern collected at 300 K, with the X-ray wavelength of 1.54 \AA. }
\end{figure}

\begin{center}
\textbf{II. Fitting parameters of constant energy scans}
\end{center}

Constant energy scans in Fig. 2 were fitted with Gaussian profiles on a slope background [Table I]:
\begin{equation}
I=y_{0}+bx+A\exp^{{-\left(x-x_{c}\right)^{2}/{2\omega^{2}}}}
\end{equation}

\noindent

\linespread{1}
\begin{table}[htbp]
\caption{Fitting parameters of constant energy scans in Fig. 2.}
\centering
\setlength{\tabcolsep}{6mm}
\large
\begin{tabular}{cccccc}
\hline \hline
\quad            & \textit{y}$_0$ & \textit{A} & \textit{b} & \textit{x}$_c$ & \textit{$\omega$} \\
\hline
SF$_x$           & 3.94(32)       & 1.55(16)   & -1.16(44)  & 0.70(1)        & 0.058(8)         \\
SF$_y$           & 3.16(23)       & 0.88(10)   & -0.78(31)  & 0.69(1)         & 0.070(10)         \\
SF$_z$           & 2.82(13)       & 0.90(21)   & -0.19(15)  & 0.69(7)         & 0.055(17)          \\
M$_y$, M$_z$     & 0.8            & 0.68(18)   & -0.41(19)  & 0.69           & 0.045(18)         \\
\hline
\end{tabular}
\end{table}

\linespread{2}


\begin{thebibliography}{}
\bibitem{Rossat-Mignod}  J. Rossat-Mignod, L. P. Regnault, C. Vettier, P. Bourges, P. Burlet, J. Bossy, J. Y. Henry, and G. Lapertot, \emph{Physica C} (Amsterdam) \textbf{185}, 86 (1991).

\bibitem{Scalapino2012RMP} D. J. Scalapino, \emph{Rev. Mod. Phys.} \textbf{84}, 1383 (2012).

\bibitem{Steglich2016RPR} F. Steglich, and S. Wirth, \emph{Rep. Prog. Phys.} \textbf{79}, 084502 (2016).

\bibitem{Matthias2006AP} M. Eschrig, \emph{Adv. Phys.} \textbf{55}, 47 (2006).

\bibitem{Dai2015RMP} P. C. Dai, \emph{Rev. Mod. Phys.} \textbf{87}, 855(2015).

\bibitem{Headings2011PRB} N. S. Headings, S. M. Hayden, J. Kulda, N. Hari Babu, and D. A. Cardwell, \emph{Phys. Rev. B} \textbf{84}, 104513 (2011).

\bibitem{Jun2011NP} J. Zhao et al., \emph{Nat. Phys.} \textbf{7}, 719 (2011).

\bibitem{Christianson2008N} A. D. Christianson et al., \emph{Nature} \textbf{456}, 930 (2008).

\bibitem{Lumsden2009PRL} M. D. Lumsden et al., \emph{Phys. Rev. Lett.} \textbf{102}, 107005 (2009).

\bibitem{Chi2009PRL} S. Chi et al., \emph{Phys. Rev. Lett.} \textbf{102}, 107006 (2009).

\bibitem{Inosov2010NP} D. S. Inosov et al., \emph{Nat. Phys.} \textbf{6}, 178 (2010).

\bibitem{Zhang2013PRL} C. Zhang et al., \emph{Phys. Rev. Lett.} \textbf{111}, 207002 (2013).

\bibitem{Zhao2013PRL} J. Zhao, C. R. Rotundu, K. Marty, M. Matsuda, Y. Zhao, C. Setty, E. Bourret-Courchesne, Jiangping Hu and R. J. Birgeneau, \emph{Phys. Rev. Lett.} \textbf{110}, 147003 (2013).

\bibitem{Steffens2013PRL} P. Steffens, et al., \emph{Phys. Rev. Lett.} \textbf{110}, 137001 (2013).

\bibitem{Lipscombe2010PRB} O. J. Lipscombe et al., \emph{Phys. Rev. B} \textbf{82}.064515 (2010).

\bibitem{Li2017PRB} Y. Li, W. Wang, Y. Song, H. Man, X. Lu, F. Bourdarot, and P. C. Dai, \emph{Phys. Rev. B} \textbf{96}, 020404 (2017).

\bibitem{WaBer2017SR} F. Wa$\rm{\beta}$er, C. H. Lee, K. Kihou, P. Steffens, K. Schmalzl, N. Qureshi, and M. Braden, \emph{Sci. Rep.} \textbf{7}, 10307 (2017).

\bibitem{Zhang2014PRB} C. L. Zhang et al., \emph{Phys. Rev. B} \textbf{90}, 140502(R) (2014).

\bibitem{Qureshi2014PRB} N. Qureshi et al., \emph{Phys. Rev. B} \textbf{90}, 100502 (2014).

\bibitem{Day2018PRL} R. P. Day et al., \emph{Phys. Rev. Lett.} \textbf{121}, 076401 (2018).

\bibitem{Daniel2018PRL} D. D. Scherer, and B. M. Andersen, \emph{Phys. Rev. Lett.} \textbf{121}, 037205 (2018).

\bibitem{Qureshi2012PRB} N. Qureshi et al., \emph{Phys. Rev. B} \textbf{86}, 060410(R) (2012).

\bibitem{npj2019}  F. Wa$\rm{\beta}$er et al., \emph{ npj Quantum Mater.} \textbf{4}, 59 (2019)

\bibitem{Ma2017PRX} M. Ma, P. Bourges, Y. Sidis, Y. Xu, S. Li, B. Hu, J. Li, F. Wang, and Y. Li, \emph{Phys. Rev. X} \textbf{7}, 021025 (2017).

\bibitem{Babkevich2011PRB} P. Babkevich et al., \emph{Phys. Rev. B} \textbf{83}, 180506(R) (2011).

\bibitem{ParkPRL2011} J. T. Park et al., \emph{Phys. Rev. Lett.} \textbf{107}, 177005 (2011).

\bibitem{Boothroyd} A. E. Taylor et al., \emph{Phys. Rev. B} \textbf{86}, 094528 (2012).

\bibitem{MWangPRB2012} M. Wang, C. Li, D. L. Abernathy, Y. Song, S. V. Carr, X. Lu,S. Li, Z. Yamani, J. Hu, T. Xiang, and P. Dai,  \emph{Phys. Rev. B} \textbf{86}, 024502 (2012).

\bibitem{inosov2012EPL} G. Friemel et al., \emph{Europhys. Lett.} \textbf{99}, 67004 (2012).

\bibitem{Wang2016PRL} Q. S. Wang et al., \emph{Phys. Rev. Lett.} \textbf{116}, 197004 (2016).



\bibitem{Pachmayr2015ACIE} U. Pachmayr, et al., \emph{Angew. Chem. Int. Ed.} \textbf{54}, 293 (2015).

\bibitem{Zhao2016NC} L. Zhao et al., \emph{Nat. Commun.} \textbf{7}, 10608 (2016).

\bibitem{Niu2015PRB} X. H. Niu et al., \emph{Phys. Rev. B} \textbf{92}, 060504 (2015).

\bibitem{Lu2015NM} X. F. Lu et al., \emph{Nat. Mater.} \textbf{14}, 325 (2015).





\bibitem{Pan2017NC} B. Pan et al., \emph{Nat. Commum.} \textbf{8}, 123 (2017).

\bibitem{Dong2015PRB} X. L. Dong et al., \emph{Phys. Rev. B} \textbf{92}, 064515 (2015).


\bibitem{Squires1978} G. L. Squires, \emph{Introduction to the Theory of Thermal Neutron Scattering} (Cambridge University Press, Cambridge, 1978).

\bibitem{Liu2012PRB} M. Liu et al., \emph{Phys. Rev. B} \textbf{85}, 214516 (2012).

\bibitem{S.V.Borisenko2016NP} S. V. Borisenko et al., \emph{Nat. Phys.} \textbf{12}, 311 (2016).

\bibitem{Liu2018PRB} D. Liu et al., \emph{Phys. Rev. B} \textbf{98}, 195137 (2018).

\end{thebibliography}
\end{document}